\documentclass[10pt,twocolumn]{article}

\usepackage{graphicx} % Command \includegraphics
\graphicspath{{figures}} % Directory to search for figures
\usepackage[font=small,labelfont=bf]{caption} % Caption control

\usepackage{xcolor}
\definecolor{black}{HTML}{212427}
\definecolor{blue}{HTML}{0563C1}
\definecolor{red}{HTML}{B51700}

\usepackage{hyperref,nameref}
\hypersetup{colorlinks=true,allcolors=blue}
\newcommand{\rref}[2]{\hyperref[#1]{\ref{#1}#2}} % Hyperref for fig. letters

\color{black}
\usepackage{setspace} % Line spacing
\usepackage[margin=0.67in]{geometry} % Margin size
\usepackage[switch]{lineno} % Line numbers (for review)
\usepackage{anyfontsize}

\usepackage{fancyhdr,lastpage}
\usepackage{titlesec}
\titleformat{\section}{\Large\bfseries}{}{0mm}{}
\titleformat{\subsection}{\bfseries}{}{0mm}{}
\titlespacing{\section}{0pt}{\baselineskip}{0pt}
\titlespacing*{\section}{0pt}{\baselineskip}{0pt}
\titlespacing*{\subsection}{0pt}{\baselineskip}{0pt}

\usepackage{amsmath,amssymb}
\newcommand{\Ang}[0]{\mathring{\mathrm{A}}} % Angstrom symbol
\renewcommand{\l}[0]{\left} % Shortcut for left modifier
\renewcommand{\r}[0]{\right} % Shortcut for right modifier
\let\f=\frac % Shortcut for fraction
\renewcommand{\t}[1]{\text{#1}} % Text mode

\usepackage{multirow} % Multi row/column in tables
\usepackage{soul} % Strike-through font
\setstcolor{red}
\setul{}{0.7pt}

\usepackage[
  backend=biber,        % Use biber
  autocite=superscript, % \autocite results in superscript, \cite is on line
  sorting=none,         % References appear in the order they are cited
  style=numeric-comp,   % Cite using numbers and compact format (e.g., 1-3)
  maxnames=100,         % Number of authors that trigger et al.
  minnames=100,         % Number of authors before et al.
  giveninits=false,     % Only initials for first name
  autopunct=false,      % Put citation before punctuation marks
  doi=true,             % Show or not DOI number
  hyperref=true         % Hyperref links in the bibliography (mostly DOI)
]{biblatex}
\DeclareFieldFormat{labelnumberwidth}{\;[#1]\!\!\!\!} % Square brackets entry
\renewbibmacro{in:}{} % Remove "in: Journal" from reference
\AtNextBibliography{\small} % Bibliography font size
\AtEveryBibitem{\clearfield{number}} % Don't show number
\AtEveryBibitem{\clearfield{month}} % Don't show month

\begin{document}

\twocolumn[
  \begin{center}
    \large
     \textbf{Nonequilibrium chemical short-range order in metallic alloys}
  \end{center}
  Mahmudul Islam$^1${\footnotemark[1]},
  Killian Sheriff$^1${\footnotemark[1]},
  Yifan Cao$^1${\footnotemark[1]}, and
  Rodrigo Freitas$^1${\footnotemark[2]} \\
  $^1$\textit{\small Department of Materials Science and Engineering, Massachusetts Institute of Technology, Cambridge, MA, USA} \\

  {\small Dated: \today}
  
  \vspace{-0.15cm}
  \begin{center}
    \textbf{Abstract}
  \end{center}
  \vspace{-0.35cm}
  Metallic alloys are routinely subjected to nonequilibrium processes during manufacturing, such as rapid solidification and thermomechanical processing. It has been suggested in the high-entropy alloy literature that chemical short-range order (SRO) could offer a new knob to tailor materials properties. While evidence of the effect of SRO on materials properties accumulates, the state of SRO evolution during alloy manufacturing remains obscure. Here, we employ high-fidelity atomistic simulations to track SRO evolution during the solidification and thermomechanical processing of alloys. Our investigation reveals that alloy processing can lead to nonequilibrium steady-states of SRO that are different from any equilibrium state. The mechanism behind nonequilibrium SRO formation is shown to be an inherent ordering bias present in nonequilibrium events. These results demonstrate that conventional manufacturing processes provide pathways for tuning SRO that lead to a broad nonequilibrium spectrum of SRO states beyond the equilibrium design space of alloys.
  \vspace{0.4cm}
]
{
  \footnotetext[1]{These authors contributed equally to this work.}
  \footnotetext[2]{Corresponding author (\texttt{rodrigof@mit.edu}).}
}

%%%%%%%%%%%%%%%%%%%%%%%%%%%%%%%%%%%%%%%%%%%%%%%%%%%%%%%%%%%%%%%
\section{Introduction}

\noindent Properties of metallic alloys are defined by their chemical composition and microstructure, i.e., the hierarchical arrangement of atoms into structures (known as defects) of different types, densities, and spatial distribution. The processing route during manufacturing is designed to lead metals to predetermined microstructures and chemical configurations. For example, after solidification the as-cast alloy is often heat treated to reduce chemical segregation, and then subjected to mechanical deformation to increase the density of linear defects (dislocations) because that strengthens the material --- a process known as work hardening. Recently, in the high-entropy materials literature\autocite{george_high-entropy_2019, schweidler_high-entropy_2024,chen_chemical_2023,schweidler2024high,han2024multifunctional}, it has been suggested that chemical short-range order (SRO) could offer a new knob to tailor materials properties. Controlling SRO during manufacturing would broaden the design space of metallic alloys without altering their microstructure and chemical composition, offering additional independent dimension of materials property design.

Chemical SRO is the preference for certain local chemical motifs over others\autocite{sheriff_quantifying_2024, sheriff2024chemicalmotif,cantor_local_2024}. This collection of chemical motifs acts as the background against which the microstructure exists, thereby affecting materials properties by modulating the behavior of crystal defects. Evidence of the effects of SRO on materials properties has been accumulating at a rapid pace, from mechanical properties\autocite{dasari_exceptional_2023,chen_simultaneously_2021} to corrosion resistance\autocite{qiu_corrosion_2017,xie_percolation_2021}, catalysis\autocite{xie_highly_2019}, and radiation-damage resistance\autocite{su_enhancing_2023,el_atwani_quinary_2023}. Controlling such materials properties through SRO requires tuning the amount of SRO present in the material, which is not an easy feat to accomplish in practice. The conventional route is to hold the alloy at a fixed temperature until thermodynamic equilibrium has been achieved (i.e., annealing), establishing with it the corresponding equilibrium SRO at that temperature\autocite{moniri_three-dimensional_2023,chen_direct_2021, zhang_short-range_2020,coury_comprehensive_2024}. Higher annealing temperatures lead to more disorder and less SRO, while lower temperatures lead to more SRO. Under this approach the lowest amount of SRO is limited by the melting temperature, which is the highest temperature the alloy can withstand in its solid phase. At intermediate temperatures, this annealing process is often exceedingly slow\autocite{coury_comprehensive_2024} and costly because equilibrium SRO is created through chemically-biased vacancy diffusion, which is a thermally-activated event; at room temperature most alloys would take centuries to achieve equilibrium.  After annealing, the alloy is quickly cooled down to room temperature (i.e., quenched), freezing in the desired degree of SRO.

Alloys are subjected to various other processes besides annealing during their manufacturing, most of which are nonequilibrium processes known to affect the chemical state of the alloy, such as rapid solidification and thermomechanical treatments\autocite{straumal_severe_2022,hou_revealing_2021,schuh_mechanical_2015,shahmir_effect_2016}. While equilibrium statistical physics and thermodynamics explain the SRO evolution during equilibrium processes, the state of SRO during alloy manufacturing remains obscure due to the absence of an equivalent physical framework for nonequilibrium processes. This reflects a lack of understanding of the fundamental mechanisms of SRO evolution during nonequilibrium processes. These mechanisms must be uncovered and quantified if the SRO knob is to be dialed for the practical co-design of materials properties alongside the traditional microstructure and chemical composition degrees of freedom.

Here, we show that nonequilibrium manufacturing processes lead to steady-state chemical short-range order (SRO) that cannot be accessed through equilibrium thermodynamics. Using large-scale atomistic simulations and a minimal physical model, we uncover that these nonequilibrium SRO states arise from inherent ordering biases during processing, such as dislocation motion and solidification dynamics. We construct a physical framework that captures this behavior and delineates a broad spectrum of SRO states beyond the equilibrium manifold, including both quasi-equilibrium and far-from-equilibrium regimes.

%%%%%%%%%%%%%%%%%%%%%%%%%%%%%%%%%%%%%%%%%%%%%%%%%%%%%%%%%%%%%%%

\section{Results}

\begin{figure*}[!tb]
  \centering
  \includegraphics[width=\textwidth]{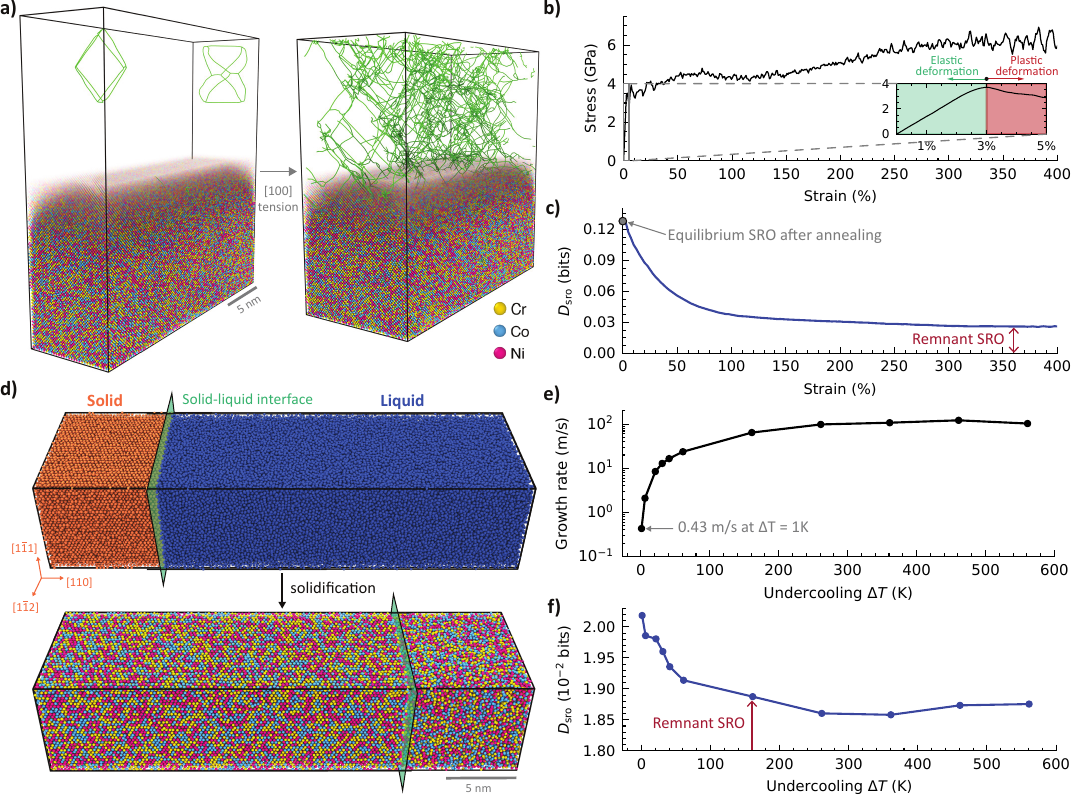}
  \caption{\label{figure_1} Remnant SRO after nonequilibrium materials processing.  \textbf{a)} CrCoNi is annealed at $1000\,\t{K}$ before undergoing uniaxial tensile deformation at room temperature ($300\,\t{K}$) under a constant strain rate of \(10^9 \t{s}^{-1}\). Dislocation network shown in green. \textbf{b)} Stress-strain curve shows strengthening plateau. \textbf{c)} Evolution of SRO ($D_\t{sro}$) reveals that mechanical deformation leads to an ultimate steady-state with finite SRO. \textbf{d)} Solidification of CrCoNi at $\Delta T = 1\,\t{K}$ undercooling ($\Delta T = T_\t{m} - T$, where $T_\t{m} = 1661\,\t{K}$ is the melting temperature). \textbf{e)} Growth rate of the solid under various undercooling temperatures. \textbf{f)} The amount of remnant SRO in the as-cast alloy has a mild dependence on the undercooling temperature. The reported $D_\t{sro}$ values of as-cast alloy correspond to the final solidified structure.}
\end{figure*}

\subsection{Materials-processing induced SRO}

Large-scale molecular dynamics simulations of solidification and thermomechanical processing of the chemically-complex CrCoNi alloy were carried out using a high-fidelity machine learning interatomic potential\autocite{cao_capturing_2024}. This potential was specifically designed to capture chemical SRO with accuracy comparable to electronic-structure calculations, including SRO effects on the solid phase, liquid phase, and microstructural elements such as dislocations. A collection of predictions made with this potential has been demonstrated to be in good agreement with experimental results.

The simulations of thermomechanical processing, illustrated in fig.~\rref{figure_1}{a}, were performed by first annealing the alloy at a high temperature, which established an initial equilibrium SRO state. The alloy was then subjected to work hardening through uniaxial tensile deformation at room temperature (300\,K), where a constant strain rate of $10^9 \, \t{s}^{-1}$ was applied along the $[100]$ crystallographic direction. The stress measured during the tensile deformation after annealing at 1000\,K is shown as a function of the total strain in fig.~\rref{figure_1}{b}, where it can be seen that the elastic deformation regime is followed by a plastic deformation regime in which strengthening occurs until a plateau is achieved\autocite{zepeda-ruiz_probing_2017,zepeda-ruiz_atomistic_2021}.

The mechanism behind the stress-strain curve in fig.~\rref{figure_1}{b} is the motion and multiplication of dislocations to accommodate the imposed strain rate, leading to the development of a complex network of interacting dislocations. When dislocations move, they shuffle atoms near their core by breaking certain chemical bonds and forming new ones; a process that has been reasonably presumed to lead to the destruction of SRO\autocite{abu-odeh_modeling_2022, yin_atomistic_2021, cao_maximum_2022,yu_origin_2015,odunuga_forced_2005}. Here we have quantified this SRO destruction using an approach based on machine learning and information theory established in ref.~\cite{sheriff_quantifying_2024} (see the Methods section \nameref{methods:LCMC}). In this approach the collection of chemical motifs observed during the simulation is first identified using machine learning, and then compared to the distribution of chemical motifs expected in a chemically random alloy. This comparison is performed using a rigorous metric of information known as Jensen-Shannon divergence\autocite{lin_divergence_1991,nielsen_generalization_2020} ($D_{\t{sro}}$). Larger values of $D_{\t{sro}}$ indicate higher amounts of SRO, while $D_{\t{sro}} = 0$ indicates a chemically random alloy (i.e., the absence of SRO). The evolution of SRO during deformation (fig.~\rref{figure_1}{c}) shows that SRO is being progressively destroyed by dislocation motion. However, we find that mechanical deformation does not lead to the complete annihilation of SRO; the SRO trajectory in fig.~\rref{figure_1}{c} indicates that dislocation motion leads to an ultimate steady-state with finite SRO, which we refer to as remnant SRO.

The plastic deformation of alloys can be rigorously regarded as a nonequilibrium process\autocite{langer_thermodynamic_2017,langer_thermodynamic_2010,averback_phase_2021,pant_phase_2021,schwen_compositional_2013,zepeda-ruiz_atomistic_2021, zepeda-ruiz_probing_2017}. The system in fig.~\rref{figure_1}{a} is driven persistently by the mechanical energy provided by the applied strain, which is continuously dissipated by dislocations as they move, multiply, and annihilate each other\autocite{langer_effective_2024}. The remnant SRO observed in fig.~\rref{figure_1}{c} is an emergent outcome of this complex system. One could reasonably expect\autocite{averback_phase_2021,pant_phase_2021,schwen_compositional_2013} remnant SRO to originate from the competition between thermally-activated vacancy diffusion, which drives SRO towards equilibrium, and chemical mixing due to dislocation motion, which has been presumed to be completely randomizing. But we find that chemical equilibration due to vacancy diffusion is negligible during mechanical deformation at room temperature under the applied strain rate. This demonstrates that the mechanism leading to remnant SRO is an inherent ordering bias during dislocation motion.

We turn now to investigate the process of solidification from the melt using the simulations illustrated in fig.~\rref{figure_1}{d}. The temperature $T$ in these simulations was held at a fixed value lower than the melting temperature\autocite{cao_capturing_2024} $T_\t{m} = 1661\,\t{K}$, which defines the degree of undercooling $\Delta T = T_\t{m} -T$. The rate at which the solid grows (i.e., the solid-liquid interface velocity in fig.~\rref{figure_1}{d}) was measured for a wide range of undercooling temperatures, as shown in fig.~\rref{figure_1}{e}. The growth-rate dependence on $\Delta T$ (fig.~\rref{figure_1}{e}) can be understood through the Wilson-Frenkel model (see, for example, ref.~\cite{freitas_uncovering_2020}) as follows. At low $\Delta T$, the growth rate is primarily limited by the thermodynamic driving force of solidification (i.e., free energy difference between the solid and liquid phases), resulting in an increase in growth rate with increasing $\Delta T$ because the driving force increases with $\Delta T$. However, as undercooling increases, the limiting factor gradually shifts from the thermodynamic driving force to the diffusion kinetics at the interface because the diffusivity decreases with $\Delta T$. Thus, at high $\Delta T$, atomic mobility is too slow to keep up with the increasing driving force, and the growth rate stagnates leading to the observed plateau in fig.~\rref{figure_1}{e}. Here we have limited $\Delta T$ to just below the turnover point where the growth rate starts decreasing with $\Delta T$, as in that regime the concept of steady-state solidification might not be applicable due to liquid-phase instabilities\autocite{men_joint_2024}. The observed growth rates varied by more than two orders of magnitude, from as slow as $0.43\, \t{m/s}$ at $\Delta T = 1\,\t K$, to faster than $100\, \t{m/s}$ for $\Delta T \geq 261\,\t{K}$. These growth rates cover the entire spectrum of manufacturing processes, from traditional casting to the rapid solidification rates of modern additive manufacturing methods. The SRO formation during solidification was quantified by measuring the amount of SRO ($D_\t{sro}$) in the solidified (or as-cast) alloy as a function of the undercooling, as shown in fig.~\rref{figure_1}{f}. We find that lower undercoolings lead to higher amounts of SRO, but even ultrafast solidification rates do not lead to the absence of SRO; instead, a finite amount of remnant SRO is observed.

The kinetics of solidification is a complex physical process in which atoms in the liquid self-organize into the crystalline solid phase --- a transition that occurs at the interface between the solid and liquid phases\autocite{freitas_uncovering_2020,sun_mechanism_2018}. The rate at which this transition happens (i.e., growth rate) determines the SRO of the as-cast state. Slower growth rates provide liquid atoms more time to self-organize and form chemical SRO in the solid phase. However, even at $\Delta T = 1\,\t{K}$ the amount of remnant SRO ($D_\t{sro} = 0.0201 \,\t{bits}$) is still significantly lower than the equilibrium SRO at the melting temperature ($D_\t{sro} = 0.0323 \,\t{bits}$). This indicates that manufactured as-cast alloys should not be considered devoid of SRO, nor considered to have the SRO of an equilibrium alloy at the melting point\autocite{han_ubiquitous_2024}. The lowest amount of remnant SRO is obtained for $\Delta T \geq 261\,\t{K}$, which coincides with the plateau in growth rate (figs.~\rref{figure_1}{e} and \rref{figure_1}{f}). Under these far-from-equilibrium conditions the liquid atoms have little time to self-organize.

\subsection{Physical framework for nonequilibrium SRO}

\begin{figure}[!tb]
  \centering
  \includegraphics[width=\columnwidth]{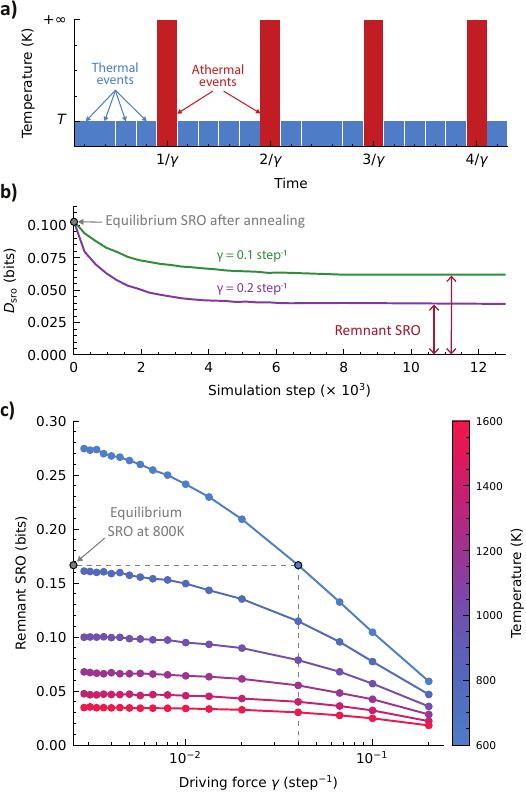}
  \caption{\label{figure_2} A simple physical model for nonequilibrium SRO. \textbf{a)} The model has only two possible dynamical events, which are enough to capture the effect of nonequilibrium processes on SRO. Thermal events guide the alloy towards equilibrium SRO at temperature $T$, while athermal events drive the alloy towards random chemical mixing. Athermal events are periodically introduced with frequency $\gamma$. \textbf{b)} Monte Carlo simulations demonstrate that this simple physical model captures the emergence of remnant SRO. \textbf{c)} Spectrum of remnant SRO as a function of the model's temperature and nonequilibrium driving force $\gamma$. The highlighted state at $600\,\t{K}$ and $\gamma = 0.04 \; \t{step}^{-1}$ has nearly the same amount of SRO ($D_\t{sro}$) as an alloy in equilibrium at $800\,\t{K}$.}
\end{figure}

Equilibrium statistical physics and thermodynamics --- which define the SRO of equilibrium processes\autocite{sheriff_quantifying_2024} --- cannot explain the remnant SRO observed during materials processing (fig.~\ref{figure_1}). A physical framework that explicitly accounts for nonequilibrium processes is needed. To derive this framework, we introduce in fig.~\rref{figure_2}{a} the simplest physical model capable of accounting for the effects of nonequilibrium processes on SRO (see the Methods section \hyperref[method:neq]{Physical model for nonequilibrium SRO}). In this model only two types of dynamical events are in operation: thermal and athermal. During thermal events a random pair of atoms is selected, and their positions are swapped following an acceptance probability given by the traditional Metropolis-Hastings algorithm\autocite{metropolis_equation_1953,hastings_monte_1970}. Such thermal events are characterized by their temperature $T$, and guide the alloy towards the corresponding equilibrium SRO at this temperature. Athermal events are similar, but the swapping of atoms is always accepted, which drives the alloy towards random chemical mixing. The role of athermal events is to mimic the forced chemical mixing of nonequilibrium processes.

Athermal events are periodically introduced in between thermal events with frequency $\gamma$, as illustrated in fig.~\rref{figure_2}{a}. The parameter $\gamma$ is the nonequilibrium driving force in this model because it quantifies the deviation from equilibrium conditions, which is a role analogous to the strain rate during mechanical deformation (fig.~\rref{figure_1}{a}) or the degree of undercooling during solidification (fig.~\rref{figure_1}{d}). The SRO evolution derived from the model (fig.~\rref{figure_2}{b}) shows that the competition between thermal and athermal events leads to a steady-state of finite SRO, which is similar to what was observed for mechanical deformation (fig.~\rref{figure_1}{c}). This demonstrates that the model captures the emergence of remnant SRO. The effect of $\gamma$ on remnant SRO is shown in fig.~\rref{figure_2}{c}, where we find that remnant SRO converges to the equilibrium SRO as $\gamma \rightarrow 0$. Meanwhile, increasing $\gamma$ drives the alloy to far-from-equilibrium conditions with remnant SRO much different from its equilibrium value.

\begin{figure}[!tb]
  \centering
  \includegraphics[width=\columnwidth]{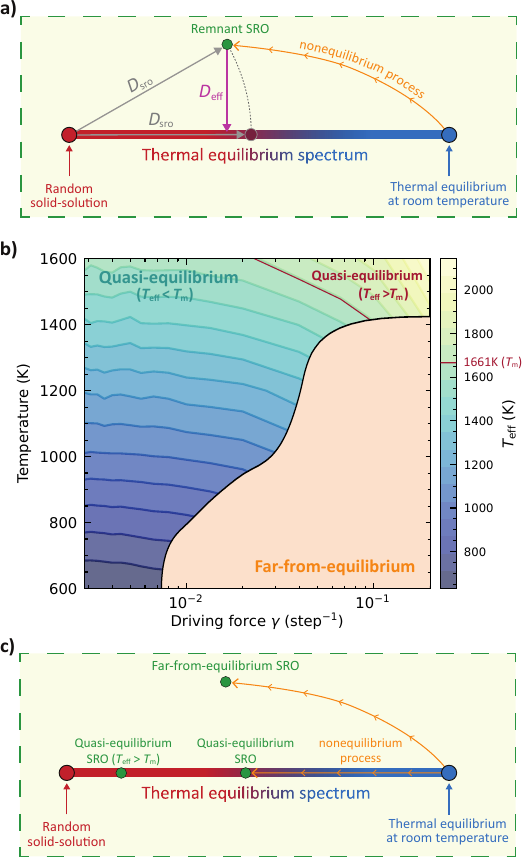}
  \caption{\label{figure_3} Physical framework for nonequilibrium SRO. \textbf{a)} The remnant SRO state (brown circle) has the same amount of SRO (i.e., $D_\t{sro}$) as the equilibrium state (dark red circle). $D_\t{eff}$ measures how far the remnant SRO is from the equilibrium spectrum of SRO states. \textbf{b)} Diagram of nonequilibrium SRO states as function of materials processing conditions (temperature and driving force). The effective temperature $T_\t{eff}$ of a remnant SRO state (indicated by the contour lines) is the temperature of the equilibrium SRO state closest to it. Quasi-equilibrium SRO states are physically indistinguishable from their effective equilibrium states. Far-from-equilibrium SRO states do not correspond to any equilibrium state. The far-from-equilibrium region (orange) comprises states whose $D_\t{eff}$ values exceed the expected $D_\t{eff}$ under equilibrium. These states do not possess a physically meaningful $T_\t{eff}$. \textbf{c)} Illustration of how nonequilibrium processes lead to the three states shown in fig.~\rref{figure_3}{b}. $T_\t{m} = 1661\,\t{K}$ is the melting temperature.}
\end{figure}

Notice in fig.~\rref{figure_2}{c} how the remnant SRO at $600\,\t{K}$ and $\gamma = 0.04 \; \t{step}^{-1}$ has nearly the same $D_\t{sro}$ as the equilibrium SRO of an alloy at $800\,\t{K}$. This raises the question of whether these two SRO states are physically equivalent. But this equivalency cannot be established by simply comparing the $D_\t{sro}$ value of both states because $D_\t{sro}$ is a metric of the amount of SRO in the alloy: it measures only the distance between the SRO state from the chemical state of a random alloy, as illustrated in fig.~\rref{figure_3}{a}. It is reasonable that these two SRO states could have the same amount of SRO (i.e., $D_\t{sro}$) while being physically distinct. We have developed a rigorous approach to establish whether or not remnant and equilibrium SRO states are equivalent, which is described next.

We start by first evaluating the Jensen-Shannon divergence between the remnant SRO and every possible equilibrium SRO state. The lowest value of the divergence identifies the effective equilibrium state of the remnant SRO, illustrated in fig.~\rref{figure_3}{a} by $D_\t{eff}$ (see the Methods section \hyperref[method:effective]{Effective temperature calculation} for more details). This equilibrium SRO state, characterized by a temperature $T_\t{eff}$, is the equilibrium state most similar to the remnant SRO of an alloy at temperature $T$ under a nonequilibrium driving force $\gamma$. Figure \rref{figure_3}{b} shows how $T_\t{eff}$ varies with $T$ and $\gamma$.

Let us consider now how to use $D_\t{eff}$ (fig.~\rref{figure_3}{a}) to establish whether or not the remnant SRO of an alloy at temperature $T$ and driving force $\gamma$ is equivalent to the equilibrium SRO state at temperature $T_\t{eff}$. Naively, one could erroneously suggest that the states are equivalent only if $D_\t{eff} = 0$. Yet even equilibrium states have $D_\t{eff} \neq 0$ due to the nature of thermal fluctuations. Hence, the correct criteria to determine equivalency is to establish whether or not $D_\t{eff}$ is statistically distinguishable from its expected value in equilibrium. This analysis identifies an equilibrium state with temperature $T_\t{eff}$ that is physically indistinguishable from the remnant SRO of an alloy at temperature $T$ under a nonequilibrium driving force $\gamma$. We call these states quasi-equilibrium SRO because the nonequilibrium process is quasistatic in the sense that linear-response theory is appropriate and the fluctuation-dissipation theorem is applicable\autocite{callen_irreversibility_1951}. In the language of the field of driven alloys we can say that the law of corresponding states proposed by Martin\autocite{martin_phase_1984} is rigorously obeyed for SRO under these driven conditions.

Finally, the states of remnant SRO with $D_\t{eff}$ outside of the equilibrium range are physically distinguishable from their corresponding effective equilibrium state at temperature $T_\t{eff}$. Consequently, the law of corresponding states\autocite{martin_phase_1984} is not an appropriate framework to describe the SRO of such states. We call these states far-from-equilibrium SRO because they can only be achieved in processes with large nonequilibrium driving forces, i.e., beyond linear-response theory approximations. Far-from-equilibrium states, illustrated in fig.~\rref{figure_3}{b}, cannot be achieved through thermal equilibration of the alloy, such as annealing processes. 

The physical framework for nonequilibrium processes introduced in figs.~\ref{figure_2} and \rref{figure_3}{a} can be summarized in a diagram of nonequilibrium SRO states, shown in fig.~\rref{figure_3}{b}. This diagram explains the effect of nonequilibrium processes on SRO (fig.~\rref{figure_3}{c}), which can be used to understand the remnant SRO observed during materials processing (fig.~\ref{figure_1}).

\subsection{Nonequilibrium SRO during materials processing} \label{section:4}

\begin{figure*}[!tb]
  \centering
  \includegraphics[width=\textwidth]{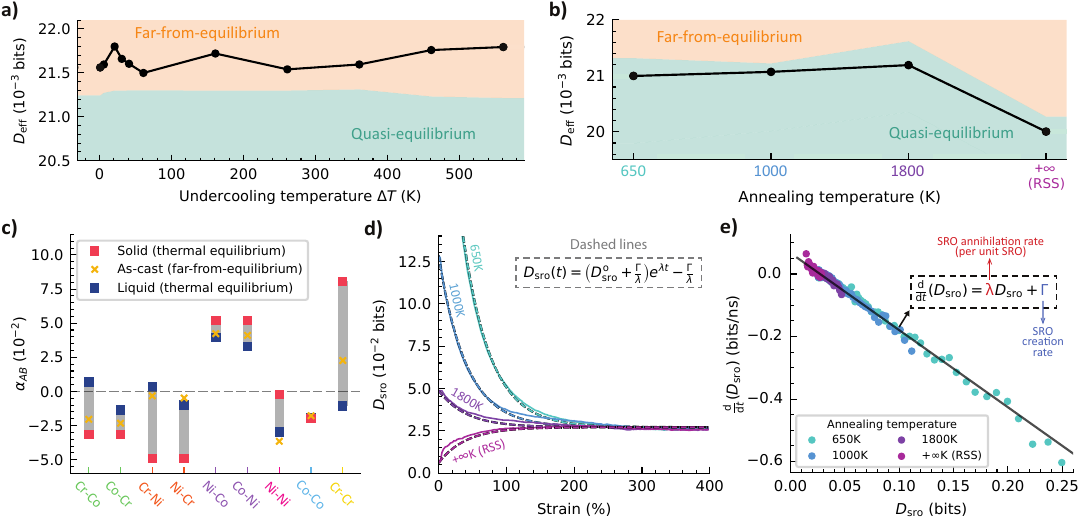}
  \caption{\label{figure_4} Nonequilibrium SRO during materials processing. The remnant SRO states created during \textbf{a)} solidification (fig.~\rref{figure_1}{d}) are far-from-equilibrium, while \textbf{b)} thermomechanical processing (fig.~\rref{figure_1}{a}) leads to quasi-equilibrium states. The reported $D_\t{eff}$ values correspond to the final snapshots of the simulations. The boundary between quasi-equilibrium (green) and far-from-equilibrium (orange) regimes is defined by calculating the expected value of $D_\t{eff}$ in equilibrium. \textbf{c)} Most of the Warren-Cowley parameters ($\alpha_{AB}$) of the as-cast alloy ($\Delta T = 1\,\t{K}$ undercooling) are closer to the liquid phase than to the solid phase, suggesting that a large portion of the far-from-equilibrium SRO is inherited from the liquid. The grey shaded region denotes the range of $\alpha_{AB}$ values between the solid and liquid equilibrium states. \textbf{d)} Evolution of the amount of SRO ($D_\t{sro}$) during mechanical deformation of samples prepared using different thermal treatments. The predictive model for $D_\t{sro}$ (inset equation and dashed lines) is derived from the observed linear behavior of the \textbf{e)} entropy-production rate on $D_\t{sro}$.}
\end{figure*}

We now turn to the analysis of the remnant SRO observed during solidification and thermomechanical processing (fig.~\ref{figure_1}) using the physical framework for nonequilibrium SRO developed in figs.~\ref{figure_2} and \ref{figure_3}. Figure \rref{figure_4}{a} shows that the remnant SRO in the as-cast alloy is far-from-equilibrium for any degree of undercooling $\Delta T$ employed during solidification. The lack of reduction in $D_\t{eff}$ with decreasing $\Delta T$ is a result that warrants a brief discussion in light of the increase in SRO with decreasing $\Delta T$ observed in fig.~\rref{figure_1}{f}. Smaller $\Delta T$ provides liquid atoms more than two orders of magnitude (fig.~\rref{figure_1}{e}) more time to self-organize and form chemical SRO, but fig.~\rref{figure_4}{a} shows that liquid atoms do not use that time to find SRO configurations closer to the ones observed in equilibrium states. This indicates that the ordering bias during solidification is such that the nature of the nonequilibrium SRO formed is fundamentally different from equilibrium SRO --- more time to self-organize results in more intense SRO of this different, far-from-equilibrium kind. Figure \rref{figure_4}{c} suggests that a large portion of the SRO of the as-cast alloy is in fact inherited from the liquid phase: notice how most of the Warren-Cowley parameters ($\alpha_\t{AB}$) of the solidified alloy are closer to the liquid phase than to the solid phase, even at the smallest undercoolings. This observation suggests the following physical mechanism of SRO formation during solidification. It is well-established that the solid-liquid interface is a region of finite width that is structurally and chemically different from both, the solid and the liquid. During solidification, atoms in the solid-liquid interface undergo a diffusion process that leads to their final arrangement into the as-cast crystalline structure. Thus, the nature of this diffusion process (which can be significantly different from the diffusion in the liquid and solid) dictates the resulting SRO state of the as-cast alloy. In ref.~\cite{freitas_uncovering_2020} it was observed that the solid-liquid interface is a liquid-like region that has its structure altered (with respect to liquid in the bulk) by the influence of the nearby solid surface --- an effect that was called interface-induced ordering. Here, we build on that observation to propose that the diffusion mechanism in the solid-liquid interface is also more akin to those of the liquid, which is substantially different than the vacancy-driven diffusion that creates equilibrium SRO in solids. As the undercooling temperature decreases, the increased time available for atomic rearrangement in the liquid phase allows for more of the chemically-biased diffusion in the solid-liquid interface to take place --- effectively equilibrating the SRO according to the structure and diffusion in the solid-liquid interface. This results in far-from-equilibrium SRO that progressively deviates from the liquid SRO as undercooling decreases, but that is still significantly different from the bulk solid state. This mechanism of SRO formation is behind recent experimental observations and empirical models\autocite{bai_short-range_2023,you_ordering_2023,han_ubiquitous_2024}.

The thermomechanical processing of the alloy leads to quasi-equilibrium SRO, as shown in fig.~\rref{figure_4}{b} for the simulation of fig.~\rref{figure_1}{a} and three additional tensile tests with samples prepared using two additional annealing temperatures and a chemically random sample (i.e., a random solid solution --- RSS). The nonequilibrium SRO diagram of fig.~\rref{figure_3}{b} indicates that the alloy must go through a continuous transition between quasi-equilibrium states during the tensile tests; to better understand the mechanisms of this evolution we look into the SRO trajectory during testing. The evolution of the amount of SRO ($D_\t{sro}$) is shown in fig.~\rref{figure_4}{d}, where it can be seen that all four samples converge to the same path-independent steady state of SRO. The creation of SRO in the RSS sample demonstrates the unequivocal existence of an ordering bias during dislocation motion, which is at odds with the predominant view in the literature\autocite{abu-odeh_modeling_2022, yin_atomistic_2021, cao_maximum_2022,yu_origin_2015, ,seol2022mechanically} that SRO is simply destroyed by dislocation motion.

Figure \rref{figure_4}{e} shows that rate of SRO change with time (equivalent to the entropy production rate in nonequilibrium systems) follows a surprisingly simple linear equation: $\t{d} D_\t{sro}/\t{d}t = \lambda D_\t{sro} + \Gamma$, with $\lambda = -2.46 \, \t{ns}^{-1}$ and $\Gamma = 6.67 \times 10^{-2}\, \t{bits/ns}$. Solving this differential equation provides a predictive model of the time evolution of $D_\t{sro}$, shown in fig.~\rref{figure_4}{d}, where it can be seen that the remnant SRO is given by $D_\t{sro}(t\rightarrow\infty) = - \Gamma/\lambda$. The physical interpretation of this behavior is clear: the rate of SRO destruction by dislocations ($\lambda D_\t{sro}$) is proportional to the instantaneous amount of SRO in the sample ($D_\t{sro}$), indicating that the reactions between dislocations and chemical motifs --- which destroy SRO --- occur more frequently when the amount of SRO in the sample is higher. Meanwhile, the rate of SRO creation by dislocations $\Gamma$ does not depend on $D_\t{sro}$, indicating that SRO is created homogeneously and continuously as dislocations move. This suggests that this inherent ordering bias is part of the natural process of dislocation motion. Dislocation motion is a process driven by the applied mechanical forces, which favor transition through pathways with low energy barriers. The selection of available energy barriers depends strongly on the local chemical motifs near the dislocation core\autocite{utt_origin_2022}, thus, dislocation motion in metallic alloys is a chemically biased process that leads to the creation of nonequilibrium SRO, a process similar to the also-chemically-biased vacancy motion that leads to the creation of equilibrium SRO.

\section{Discussion}

The diagram of nonequilibrium SRO as function of materials processing conditions (fig.~\rref{figure_3}{b}) has various features that warrant further discussion within the context of alloy manufacturing. For example, the contour lines within the quasi-equilibrium region indicate that any given effective temperature $T_\t{eff}$ can be obtained under a variety of different processing conditions (i.e., different combinations of driving force and temperature). This trade-off between $T$ and $\gamma$ shows that different processing conditions can lead to the same quasi-equilibrium SRO state; this observation can be used to optimize materials processing routes for the co-design of SRO alongside the traditional microstructure and chemical composition degrees of freedom. Note also that the quasi-equilibrium region shows that it is possible to achieve effective SRO temperatures above the melting point of the alloy, i.e., $T_\t{eff} > T_\t{m}$, as demonstrated in fig.~\rref{figure_3}{b}. These states, which are rigorously characterized as equilibrium states, cannot be achieved without nonequilibrium processes because any practical equilibrium processing route would melt the alloy before reaching temperatures above $T_\t{m}$. Finally, note that the far-from-equilibrium region in fig.~\rref{figure_3}{b} shows that far-from-equilibrium SRO states can be achieved at much smaller driving forces if the nonequilibrium process is carried out at a lower temperature. This indicates that manufacturing processes operating at lower temperatures might be better suited to achieve nonequilibrium SRO --- given that controlling temperature is often easier than applying very large driving forces.

Nonequilibrium SRO can arise and be controlled by many other manufacturing processes and conditions in addition to the ones considered here (figs.~\rref{figure_1}{a} and \rref{figure_1}{d}). While these are too numerous to be accounted for in this work, we offer a brief discussion of cases just beyond the limits of our results. The most immediate of such cases is probably the reduction of the strain rate during mechanical deformation, which would lead to higher remnant SRO and different quasi-equilibrium SRO states, as suggested by figs.~\rref{figure_2}{c} and \rref{figure_3}{b}. Lower strain rates also increase the timescale of the tensile test; if the timescale is long enough for vacancy diffusion to be active then a qualitatively different mechanism for inducing order will also be active. Similar effects could be achieved by increasing the deformation temperature, which was kept at room temperature in our tensile tests. The overall effect of lowering the strain rate on SRO is not trivial, for example, in addition to vacancy activation the lower strain rate would also result in more planar slip of dislocations, which has non-obvious consequences on SRO. These examples make it clear that the parameters $\Gamma$ and $\lambda$ defining SRO evolution (fig.~\rref{figure_4}{d}) are dependent at least on the processing temperature and strain rate. Meanwhile, the timescale of the solidification conditions considered here (fig.~\rref{figure_1}{e}) covers a wide spectrum of manufacturing processes. Effects immediately beyond those considered here would certainly be those related to longer length scales\autocite{asta_solidification_2009}, such as for example, suppressing long-range chemical diffusion during ultra-fast solidification, leading to the extended solubility of chemical elements, and the associated change in the composition available for SRO formation.

Chemical and structural SRO have been extensively studied in metallic glass systems\autocite{hayes_short-range_1978, yavari_new_2006}. It is therefore instructive to consider potential parallels with the present work. In metallic glasses, chemical SRO develops within a structurally amorphous matrix that becomes frozen due to kinetic constraints, long before structural equilibration (i.e., crystallization) is achieved. These SRO states are thus inherently nonequilibrium. In contrast, the nonequilibrium SRO states investigated here emerge within crystalline solid solutions, where the underlying lattice remains structurally ordered. In these systems, the nonequilibrium character pertains solely to the chemical configuration, not to the atomic arrangement of the lattice. Consequently, the chemical SRO states examined in this work define a fundamentally distinct regime --- one in which chemical order alone evolves out of equilibrium in a structurally ordered lattice.

The physical framework for nonequilibrium SRO derived here is broadly applicable to crystalline solid solutions across various crystal structures, alloy chemistries, and manufacturing processes. It is situated within the field of so-called driven alloys. Foundational work in this field has focused on understanding the phase evolution of alloys under irradiation and severe plastic deformation\autocite{martin_driven_1996, lund_driven_2003, averback_phase_2021}, including the observation of far-from-equilibrium effects such as compositional patterning\autocite{averback_phase_2021} that cannot be captured by effective temperature models\autocite{martin_driven_1996}. The work presented here extends the understanding of SRO in driven alloys, and culminates in the nonequilibrium SRO diagram of fig.~\rref{figure_3}{b}.

The present results uncover an extensive variety of nonequilibrium SRO states, and demonstrate that SRO should not be considered a one-dimensional property lying along the random-to-thermal-equilibrium line (fig.~\rref{figure_3}{c}). The traditional manufacturing strategy used to create and control SRO (i.e., annealing) accesses only a small fraction of the possible SRO states. Our results also suggest that advanced manufacturing techniques might be particularly well-suited to explore additional regions of the SRO diagram (fig.~\rref{figure_3}{b}); for example, nonequilibrium SRO can be created through severe plastic deformation, which can be achieved by several deformation-based metal additive manufacturing techniques such as friction stir and cold spray deposition\autocite{yu_non-beam-based_2018, yin_deposition_2019}. Moreover, the slow chemical diffusion during equilibrium processes indicates that nonequilibrium processing might be the only practical route to explore the SRO diagram (fig.~\rref{figure_3}{b}). The observation of remnant SRO (fig.~\rref{figure_1}{c}) is also of interest for the field of materials under extreme conditions, where remnant SRO represents the lower bound for chemical mixing, which affects chemical reactivity in high-energy materials. The extent to which nonequilibrium SRO can be induced is strongly influenced by the chemical complexity of the alloy system. Equiatomic, multi-principal element alloys provide a richer configurational landscape\autocite{zhang_roadmap_2025}, that can potentially give rise to more diverse nonequilibrium SRO states. Other intrinsic factors\autocite{george_high-entropy_2019, schweidler_high-entropy_2024,chen_chemical-affinity_2021} such as atomic size mismatch, electronegativity differences, and mixing enthalpy may further shape the SRO diagram of fig.~\rref{figure_3}{b}, offering additional considerations for selecting alloy chemistries while exploring nonequilibrium SRO.

In summary, our work investigates the evolution of nonequilibrium SRO in crystalline alloys through large-scale atomistic simulations of traditional manufacturing processes. We derive a physical framework grounded in machine learning and information theory to characterize these states and distinguish them from their equilibrium counterparts. The results obtained in this study reveal that a broader spectrum of SRO states can be accessed only through nonequilibrium processing routes, beyond those achievable by traditional annealing. These findings thus call for alternative modes of thinking about SRO during manufacturing; in particular, it calls for strategies for materials properties optimization that co-design SRO alongside the traditional degrees of freedom --- i.e., microstructure and chemical composition.

%%%%%%%%%%%%%
% METHODS
%%%%%%%%%%%%%
\clearpage
\twocolumn[
  \begin{center}
    \Large
    \textbf{Methods}
  \end{center}
]

\subsection{Interatomic potential}
\label{method:mliap}
All simulations performed here employed the machine learning interatomic potential for CrCoNi developed in ref.~\cite{cao_capturing_2024} (training set TS-f), which is a moment tensor potential\autocite{shapeev_moment_2016}. This is a high-fidelity potential specifically designed to capture chemical SRO with accuracy comparable to density-functional theory calculations, including SRO effects on the solid phase, liquid phase, and microstructural elements such as dislocations. A collection of predictions made using this potential has been demonstrated to be in good agreement with experimental results. All simulations were carried out using the Large-scale Atomic/Molecular Massively Parallel Simulator (LAMMPS, 27 June 2024 release)\autocite{thompson_lammps_2022} to evaluate the energy and forces of this potential.

\subsection{Mechanical deformation}
\label{method:pd}
Molecular dynamics simulations of uniaxial tensile deformation were conducted on the face centered cubic (fcc) phase of equiatomic CrCoNi with a timestep of 5\,fs. The samples were prepared in four initial conditions: one random solid solution and three annealed (i.e., thermally equilibrated) samples at 650\,K, 1000\,K, and 1800\,K. The initial dimensions of the samples were $30a_0 \times 100a_0 \times 100a_0$ aligned along the [100], [010], and [001] crystallographic axes, respectively, where $a_0$ is the lattice parameter at the appropriate temperature. This resulted in approximately $1.2$ million atoms in each sample. Periodic boundary conditions were applied in all three dimensions.

After annealing, all samples were seeded with six rhombus-shaped dislocation loops created by embedding vacancy-type prismatic loops in the sample. Each dislocation loop had as Burgers vector one of the six possible $\f{1}{2}\langle 110 \rangle$ vectors. The edges of the loops were aligned along $\langle 112 \rangle$ directions (i.e., perpendicular to the Burgers vectors), situating all dislocation loops on $\{111\}$ glide planes. The rhombus edges measured $40\, \Ang$, resulting in an initial dislocation density of $1.15 \times 10^{16} \, \t{m}^{-2}$. The dislocation loops were placed randomly in the sample while avoiding any overlaps.

Following the creation of the dislocation loops, the samples were equilibrated at room temperature (300\,K) and zero pressure for $100\,\t{ps}$ using a Nosé-Hoover thermostat (damping parameter of 2\,ps and chain length of 3) and Nosé-Hoover barostat (damping parameter of 10\,ps and chain length of 3) allowing only for isotropic deformations. The sample was then mechanically deformed\autocite{zepeda-ruiz_atomistic_2021,zepeda-ruiz_probing_2017} in tension along the [100] direction at a true strain rate of $10^9 \, \t{s}^{-1} $ for a total of $4\,\t{ns}$. During deformation, the sample was allowed to contract freely along the directions orthogonal to the tensile axis (i.e., [010] and [001]); this was achieved by employing a Nosé-Hoover barostat (damping parameter of 10\,ps and chain length of 3) to maintain a total pressure of zero along the directions orthogonal to the tensile axis. The heat generated by dislocation motion was dissipated by employing a Nosé-Hoover thermostat (damping parameter of 2\,ps and chain length of 3) to maintain the temperature at 300\,K.

The dislocations densities were computed using the dislocation extraction algorithm \autocite{stukowski_extracting_2010} (DXA). All the atomistic simulation visualizations were done using Open Visualization Tool \autocite{stukowski_visualization_2009} (OVITO, version 3.13.0).

\subsection{Solidification}
\label{method:solid}
Molecular dynamics simulations of CrCoNi solidification were carried out at an equiatomic concentration and a total of 196,608 atoms (timestep of $2\,\t{fs}$). The initial configuration contained 25\% of its atoms in a cubic fcc slab with orthogonal axis aligned with the $[1\bar{1}2]$, $[1\bar{1}\bar{1}]$, and $[110]$ directions; the remaining $75\%$ of atoms are randomly distributed (i.e., liquid phase) above and below the (100) surfaces of the fcc slab. Periodic boundary conditions were applied in all directions. The initial dimensions were adjusted to accommodate for the lattice constant and liquid density at the corresponding simulation temperature, resulting in dimensions of approximately $71\,\Ang \times 100\,\Ang \times 337\,\Ang$. 

This solid--liquid system is equilibrated as follows. First, the liquid atoms coordinates were relaxed for 100 steps using the conjugate gradient method while keeping the solid atoms fixed, eliminating any atomic overlaps obtained during the creation of the liquid phase. This is followed by $2\,\t{ps}$ of equilibration of the liquid atoms with a Nosé–Hoover thermostat (damping parameter of 0.1\,ps and chain length of 3) at the target temperature while keeping the solid atoms fixed. Next, the liquid atoms are kept fixed while the solid atoms are equilibrated for $2\,\t{ps}$ at 95\% of the target temperature. Finally, the liquid atoms are equilibrated (while keeping the solid atoms fixed) for $2\,\t{ps}$ at the target temperature with a Nosé-Hoover barostat (damping parameter of 1\,ps and chain length of 3) to achieve zero pressure along the directions perpendicular to the solid--liquid interface.

After equilibration, the solidification simulations were carried out with a Nosé–Hoover thermostat with a large damping parameter of 1\,ps in order to not affect the kinetics of liquid diffusion (chain length of 3). Notice that the temperature was calculated after removing the center-of-mass velocity to avoid bias. A Nosé-Hoover barostat (damping parameter of 1\,ps and chain length of 3) was also employed to maintain zero pressure along the direction perpendicular to the solid--liquid interface, while the directions parallel to the interface are kept constant during solidification. Solidification simulations were performed from 1660\,K to 1100\,K, which correspond to undercooling temperatures ranging from $\Delta T = 1\,\t{K}$ to $\Delta T = 561\,\t{K}$ (the melting temperature was determined to be $T_\t{m} = 1661\,\t{K}$ in ref.~\cite{cao_capturing_2024}). The solidification is carried out until the system is 80\% solidified, which took from $\approx 200\,\t{ps}$ at the maximum undercooling to $\approx 35\,\t{ns}$ at $\Delta T = 1\,\t{K}$ undercooling. The total linear momentum of the center-of-mass was set to zero every $2\,\t{ps}$ to facilitate the visualization of the simulation --- note that this has no physical effect on the simulation.

The growth rate is measured by estimating the rate at which liquid atoms transform into solid atoms using the polyhedral template matching algorithm\autocite{larsen_robust_2016, stukowski_visualization_2009} with a root-mean-square deviation cutoff of 0.15. The growth rate was evaluated with data collected only after an initial equilibration period of $100\,\t{ps}$, and also only before the system is 75\% solidified in order to avoid finite system sizes.

\subsection{Equilibrium Monte Carlo simulations}
\label{method:eq}
Monte Carlo simulations were performed to thermally equilibrate the chemical configuration in equiatomic CrCoNi for temperatures ranging from $400\,\t{K}$ to $3000\,\t{K}$ in increments of $25\,\t{K}$, and from $3000\,\t{K}$ to $7000\,\t{K}$ in increments of $125\,\t{K}$. A total of 50 independent Monte Carlo simulations were performed for each temperature, where each simulation contained 4,000 atoms in a cubic fcc system with dimensions $10a_0 \times 10a_0 \times 10a_0$ oriented along the $[100]$, $[010]$, and $[001]$ crystallographic directions, where $a_0$ is the lattice parameter at the appropriate temperature. Beyond the melting point, the lattice parameters were determined by extrapolating the linear relationship between lattice parameter and temperature. Periodic boundary conditions were applied in all directions.

The simulations were initialized by randomly distributing the chemical elements in equiatomic proportions on the fcc lattice. Thermal equilibration was then achieved by performing atom-swap attempts between atoms of different chemical types, with the acceptance of swaps determined by the Metropolis criterion \autocite{metropolis_equation_1953,hastings_monte_1970} probability $\exp(-\Delta E/k_\t{B}T)$, where $k_\t{B}$ is the Boltzmann constant, $T$ is the system temperature, and $\Delta E$ is the change in energy caused by the swap. A total of 160,000 Monte Carlo steps (i.e., atom-swap attempts) were performed, or 40 steps per atom on average. Data for the statistical analyses were collected only after an initial equilibration time of 80,000 steps. Statistically uncorrelated configurations were collected every 13,250 steps during the remaining 80,000 steps, resulting in seven snapshots being collected per Monte Carlo simulation.

\subsection{Physical model for nonequilibrium SRO}
\label{method:neq}
The physical model for nonequilibrium SRO introduced in fig.~\rref{figure_2}{a} has only two types of dynamical events in operation: thermal and athermal. During thermal events a random pair of atoms is selected, and their positions are swapped following an acceptance probability given by the traditional Metropolis-Hastings algorithm\autocite{metropolis_equation_1953,hastings_monte_1970}. Such thermal events are characterized by their temperature $T$, and guide the alloy towards the corresponding equilibrium SRO at this temperature. Athermal events are similar, but the swapping of atoms is always accepted, which drives the alloy towards random chemical mixing. Athermal events are periodically introduced in between thermal events with frequency $\gamma$, as illustrated in fig.~\rref{figure_2}{a}.

In order to solve this model (i.e., obtain the results in figs.~\ref{figure_2} and ~\ref{figure_3}) we have employed the Monte Carlo method with the same machine learning potential used in the thermomechanical and solidification simulations; we refer to these simulations as nonequilibrium Monte Carlo. These simulations are nearly identical to the equilibrium Monte Carlo simulations described in the last section, with the only difference being that the atom-swap attempt of every $\gamma^{-1}$th step is always accepted, as illustrated in fig.~\rref{figure_2}{a}. This is equivalent to setting $T \rightarrow +\infty\,\t{K}$ in the Metropolis criteria, which causes the probability $\exp(-\Delta E/k_\t{B}T) \rightarrow 1$. These $T \rightarrow +\infty\,\t{K}$ swaps are athermal events that mimic the forced chemical mixing of nonequilibrium processes. 

The nonequilibrium Monte Carlo simulations were carried out using the same size, number of steps, and number of independent simulations as the equilibrium Monte Carlo simulations. They were performed at temperatures ranging from $600\,\t{K}$ to $1600\,\t{K}$, at $200\,\t{K}$ intervals. A total of 17 different values were chosen for the driving force $\gamma$, varying from $\gamma^{-1} = 5$ to $\gamma^{-1} = 350$.

\subsection{SRO quantification}
\label{methods:LCMC}
Chemical SRO quantification was performed using the chemical-motif approach introduced in refs.~\cite{sheriff_quantifying_2024} and \cite{sheriff2024chemicalmotif}. In this approach the chemical environment surrounding an atom is described by a chemical motif $\mathcal{M}_i$ consisting of the atom itself and its nearest neighbors (i.e., first coordination shell). In ref.~\cite{sheriff_quantifying_2024} it was demonstrated that this approach completely quantifies SRO up to the first coordination shell of each atom, while the traditional Warren-Cowley parameters\autocite{cowley_approximate_1950}, which focus on pairwise correlations, do not distinguish among certain local motifs. We have employed the Euclidean graph neural network\autocite{smidt_euclidean_2021} architecture provided in ref.~\cite{sheriff2024chemicalmotif} to identify all chemical motifs $\mathcal{M}_i$ in the simulations, and evaluate their population density distributions $P(\mathcal{M}_i)$. We will omit the population density distribution $P$ dependency on $\mathcal{M}_i$ for brevity.

Chemical SRO causes certain chemical motifs $\mathcal{M}_i$ to be more common than others, thereby affecting $P$. This tendency was quantified in refs.~\cite{sheriff_quantifying_2024} and \cite{sheriff2024chemicalmotif} using the Kullback–Leibler divergence\autocite{mackay2003information} with respect to the population density distribution of a random alloy $P_\t{rss}$ (i.e., an alloy without any SRO):
\[
  D_\t{KL} (P \parallel P_\t{rss}) = \sum_i P(\mathcal{M}_i) \log_2 \l[ \f{P(\mathcal{M}_i)}{P_\t{rss}(\mathcal{M}_i)} \r],
\]
where the sum is over all possible chemical motifs of the alloy. The work performed here required us to employ a different --- but similar --- quantity known as the Jensen-Shannon divergence\autocite{lin_divergence_1991,nielsen_generalization_2020}, which is more appropriate for our goals. The Jensen-Shannon divergence between two probability distributions $P_1$ and $P_2$ is defined as
\begin{equation}
    \label{eq:js}
    D_\t{JS}(P_1 \parallel P_2) = x_1 \, D_{\t{KL}}(P_1 \parallel M) + x_2 \, D_{\t{KL}}(P_2 \parallel M),
\end{equation}
where $x_n = N_n / (N_1+N_2)$ is the weight coefficient of the sample size $N_n$ of $P_n$, and $M = x_1 P_1 + x_2 P_2$ is a weighted average of the two distributions. The Jensen-Shannon divergence $D_\t{JS}$ significantly reduces the bias introduced by having samples $P_1$ and $P_2$ of different sizes\autocite{pan_novel_2022,ulger_fine-grained_2023}, which occurs often in thermomechanical deformation or solidification simulations (see the discussion below on how the motifs are extracted during these simulations). The Jensen-Shannon divergence also has other advantages over the Kullback–Leibler divergence, such as: it is symmetric with respect to $P_1$ and $P_2$, it is bounded, it can be applied to distributions with zero counts, and $\sqrt{D_\t{JS}}$ is a rigorous mathematical metric\autocite{endres_new_2003} --- which makes $D_\t{JS}$ more appropriate to quantify the amount of SRO than $D_\t{KL}$. The amount of chemical SRO in a distribution $P_1 = P$ is quantified by setting $P_2 = P_\t{rss}$ in eq.~\ref{eq:js}:
\begin{equation}
  \label{eq:D_sro}
  D_\t{sro} = D_\t{JS}(P \parallel P_\t{rss}).
\end{equation}

For the equilibrium and nonequilibrium Monte Carlo simulations the chemical-motif population density $P$ is evaluated by considering the data collected from all seven snapshots of each of the 50 independent simulations (i.e., $7 \times 50 \times 4,000 = 1,400,000$ motifs for each case). This is done for each temperature, and for each value of driving force $\gamma$ in the case of nonequilibrium simulations.

The evaluation of $P$ resulting from solidification is more complex. First, the atoms in the initial crystalline seed of the final solidified structures are excluded from the calculation because their chemical motifs are not resulting from the solidification process. The last 25\% of the atoms to solidify are also excluded to avoid the influence of finite-size effects (similar to what was done in the evaluation of the growth rate). Finally, the as-cast alloy contained vacancies, which disrupted the identification of chemical motifs. To avoid this problem the crystal structure surrounding the remaining atoms was classified using the polyhedral template matching (PTM) algorithm\autocite{larsen_robust_2016} (with a root-mean-square deviation cutoff parameter of 0.15), and only atoms classified as having a fcc structure were used in the evaluation of $P$.

The mechanical deformation simulations also required the use of PTM as a pre-processing step because the identification of chemical motifs near dislocation cores is ambiguous. In this case it is necessary to consider two crystal structures --- namely, fcc and hexagonal close-packed (hcp) --- because the decomposition of $\f{1}{2}\langle 110 \rangle$ dislocations into Shockley partial dislocations $\f{1}{6} \langle 112 \rangle$ leads to the formation of stacking faults in which the local crystal structure (up to the first coordination shell) is hcp, which has chemical motifs that are topologically different from the ones in the fcc structure. In order to accommodate both types of chemical motifs the amount of SRO is quantified as
\[
  D_\t{sro} = f_\t{fcc} \, D_\t{JS}\l(P_\t{fcc} \parallel P_\t{rss}^\t{fcc}\r) + \xi f_\t{hcp} \, D_\t{JS}\l(P_\t{hcp} \parallel P_\t{rss}^\t{hcp}\r),
\]
where $f_\alpha$ is the fraction of the system in the $\alpha$ crystal structure ($\alpha =$ fcc or hcp), $P_\alpha$ is the chemical-motif population density of $\alpha$, $P^\alpha_\t{rss}$ is the population density of a random alloy in the crystal structure $\alpha$, and $\xi = \log_2(36,333) / \log_2(140,022)$ is a counting parameter that ensures that the maximum possible contribution of both crystal structures are equal in magnitude (36,333 is the total number of possible motifs in a ternary fcc crystal, and 140,022 is the equivalent quantity for hcp).

\subsection{Annealing simulations}
\label{method:rmc}

Annealing the samples for the mechanical deformation simulations with direct Monte Carlo is unfeasible due to the prohibitive computational cost of this method: the annealing would cost nearly one order of magnitude more than all other simulations in this work combined. To circumvent this, a variation of the reverse Monte Carlo method was employed to produce samples with appropriate equilibrium SRO.

The reverse Monte Carlo method has two objective functions; the first one is
\[ 
  \chi_\t{WC}^2 = \sum_{AB} \l(\alpha_{AB}^\t{target} - \alpha_{AB}\r)^2,
\]
where the sum is over all chemical element pairs ($A$ and $B$), and
\begin{equation}
  \label{eq:wc}
  \alpha_{AB} = 1 - \f{P(A|B)}{c_{A}}
\end{equation}
is the Warren-Cowley parameter\autocite{cowley_approximate_1950} of the $AB$ pair where $p(A|B)$ is the conditional probability of finding $A$ at a nearest-neighbor site of $B$, and $c_{A}$ is the average concentration of $A$ in the alloy. This objective function measures the deviation of the sample's Warren-Cowley parameters from their target values $\alpha_{AB}^{\t{target}}$, which are the equilibrium Warren-Cowley parameters at the annealing temperature. The second objective function is
\[ 
  \chi_\t{sro}^2 = (D_\t{sro}^\t{target} - D_\t{sro})^2,
\]
where $D_\t{sro}^\t{target}$ is the equilibrium $D_\t{sro}$ (eq.~\ref{eq:D_sro}) at the annealing temperature.

The objective functions are minimized using a simulated annealing approach with the Metropolis-Hastings algorithm. Each atom-swap step $n$ of the simulation is divided into two parts, first we evaluate whether the swap satisfies the first objective function with acceptance probability given by
\[
  \exp\l( -\beta_{n+1} \, \Delta \chi_\t{WC}^2 \r),
\]
where $\beta_{n+1} = (1 + n) \beta_n$ (with $\beta_0 = 10$), and $\Delta \chi_\t{WC}^2$ is the change in the objective function caused by the swap. The atom-swap is rejected if this first part fails, but if it succeeds then the algorithm proceeds to test whether the swap satisfies the second objective function, with acceptance probability given by
\[
  \exp\l( -\beta_{n+1} \, \Delta \chi_\t{sro}^2 \r),
\]
where $\Delta \chi_\t{sro}^2$ is the change in the objective function caused by the swap. If this part succeeds the atom-swap move is accepted.

This approach is much more computationally efficient as it avoids the evaluation of atomic interactions using the machine learning interatomic potential, and it minimizes the evaluation of $D_\t{sro}$, both of which are computationally expensive.

\subsection{Effective-temperature calculation}
\label{method:effective}

The effective temperature $T_\t{eff}$ of a nonequilibrium SRO state is defined as the temperature $T$ of the equilibrium SRO state that is closest to it in terms of chemical-motif $\mathcal{M}_i$ population density. In practice this is evaluated by first computing the Jensen-Shannon divergence between the nonequilibrium SRO state $P$ and all the possible equilibrium states $P_\t{eq}(T)$, and then identifying the minimum of $D_\t{JS}(P \parallel P_\t{eq}(T))$ as function of $T$. Thus, the effective temperature can be expressed as 
\[
  T_\t{eff} = T \Big|_{\min \left[ D_{\t{JS}}(P \parallel P_\t{eq}(T)) \right]},
\]
where $D_\t{eff} = \min \left[ D_{\t{JS}}(P \parallel P_\t{eq}(T))\right]$ measures how far the nonequilibrium SRO state is from its effective equilibrium SRO counterpart (fig.~\rref{figure_3}{a}).

The criterion for determining equivalency between a nonequilibrium state and its corresponding equilibrium SRO state is to establish whether or not its $D_\t{eff}$ is statistically distinguishable from its expected value in equilibrium. This is used to classify the remnant SRO states as quasi-equilibrium or far-from-equilibrium in fig.~\rref{figure_3}{b}.

\subsection{Liquid Warren-Cowley parameters}

The Warren-Cowley parameters of the solid phase and as-cast state in fig.~\rref{figure_4}{c} are evaluated using eq.~\ref{eq:wc}, which is a traditional approach in the literature. This approach is not applicable to the liquid phase. Instead, the Warren-Cowley for the liquid were evaluated following the analysis from refs.~\cite{asta1999embedded} and \cite{maret1990structure}, which defined
\[
  \alpha^\t{liquid}_{AB} = 1 - \f{z_{AB}}{c_{B} Z_{A}},
\]
where $z_{AB}$ is the average number of atoms of type $B$ surrounding atoms of type $A$ in the first coordination shell, and $Z_{A} = \sum_B z_{AB}$ is the coordination number around atoms $A$. We computed $z_{AB}$ by integrating the partial radial distribution function $g_{AB}(r)$ up to the cutoff distance $r_\t{cut}$ for the first coordination shell,
\[
  z_{AB} = \int_{0}^{r_\t{cut}} \rho \, g_{AB}(r) \, 4\pi r^2 \, \t{d}r,
\]
where $\rho$ is the average number density of atoms in the liquid, and $r_\t{cut}$ is determined as the first minima in the total radial distribution function $g(r)$. Note that the liquid WC parameters are not symmetric due to the different coordination number of different chemical elements, i.e., in general $\alpha^\t{liquid}_{AB} \neq \alpha^\t{liquid}_{BA}$, as shown in fig.~\rref{figure_4}{c}.

\subsection{Data and code availability}
        
The software for chemical motif identification and SRO quantification can be found in our ChemicalMotifIdentifier Python package\autocite{sheriff_killiansheriffchemicalmotifidentifier_2025}. The machine learning potential can be found in our MachineLearningPotential GitHub repository\autocite{potentialgithub}. Our figure style is implemented in LovelyPlots\autocite{lovelyplots} under the paper style. Any custom code that is not currently available can be subsequently added upon request to the corresponding author.

\clearpage
\printbibliography[heading=bibnumbered,title={References}]

\clearpage

\subsection{Acknowledgments}

This work was supported by the MathWorks Ignition Fund, MathWorks Engineering Fellowship Fund, and the Portuguese Foundation for International Cooperation in Science, Technology and Higher Education in the MIT--Portugal Program. This material is based upon work supported by the Air Force Office of Scientific Research (AFOSR) under award number under award number FA9550-25-1-0199, through the Young Investigator Program. We were also supported by the Research Support Committee Funds from the School of Engineering at the Massachusetts Institute of Technology. This work used the Expanse supercomputer at the San Diego Supercomputer Center through allocation MAT210005 from the Advanced Cyberinfrastructure Coordination Ecosystem: Services \& Support (ACCESS) program, which is supported by National Science Foundation grants \#2138259, \#2138286, \#2138307, \#2137603, and \#2138296, and the Extreme Science and Engineering Discovery Environment (XSEDE), which was supported by National Science Foundation grant number \#1548562.

\subsection{Author contributions statement}

M.I., K.S., Y.C., and R.F. conceived the project.
M.I. developed the nonequilibrium SRO model and performed effective temperature calculations.
M.I. conducted Monte Carlo and mechanical deformation simulations.
K.S., and M.I. quantified SRO. 
K.S. conducted the annealing simulations. 
Y.C. conducted the solidification simulations and quantified liquid Warren-Cowley parameters.
All authors contributed to the interpretation of the results, prepared, reviewed, and edited the manuscript.
Project administration and supervision were performed by R.F.

\subsection{Competing interests statement}

The authors declare no competing interests.

\end{document}